\documentstyle[psfig]{mn}
\def\sa{supernova\/}
\def\se{supernov\ae\/}

\def\ltsima{$\; \buildrel < \over \sim \;$}
\def\simlt{\lower.5ex\hbox{\ltsima}}
\def\gtsima{$\; \buildrel > \over \sim \;$}

\def\simgt{\lower.5ex\hbox{\gtsima}}            

\def\hii{{\sc Hii\/}}

\def\ha{{\sc H}$\alpha$}

\def\cm2{cm$^{-2}$\/}
\def\cm3{cm$^{-3}$\/}

\def\o4363{{\sc{[Oiii]}}$\lambda$4363\/}

\def\f{{$\mathcal{S}$}}
\def\kms{km~s$^{-1}$}
\begin{document}
\title{On Core Collapse Supernov\ae\ in Normal and in Seyfert Galaxies}
\author[A Bressan et al.]{A. Bressan$^{1}$, M. Della Valle$^{2,3}$, P. Marziani$^{1}$\\
$^1$ Osservatorio Astronomico di Padova, Vicolo dell' Osservatorio
5, I-35122 Padova, Italy \\
$^2$ Osservatorio Astrofisico di Arcetri, Firenze, Italy\\
$^3$ European Southern Observatory, Garching by M\"unchen, Germany }
\date{Received: never; Accepted: never}

\maketitle

\label{firstpage}
\begin{abstract}
This paper  estimates the relative frequency of different types of
core-collapse \se, in terms of the ratio  \f\ between the number of
type Ib--Ic and  of type II \se. We estimate \f\ independently for
all normal and Seyfert galaxies whose   radial velocity  is
$\le$14000 \kms, and which had at least one \sa\ event recorded in
the Asiago catalogue from January 1986 to August 2000. We find that
the ratio \f\ is $\approx$ 0.23$\pm$0.05 in normal galaxies. This
value is consistent with constant star formation rate and with a
Salpeter Initial Mass Function  and average binary rate $\approx$
50\%. On the contrary, Seyfert galaxies exceed the ratio \f\ in
normal galaxies by a factor $\approx$ 4  at a confidence level $\ga 2
\sigma$. A caveat is that the numbers for Seyferts are still small
(6 type Ib-Ic and 6 type II supernovae discovered as yet). Assumed
real, this excess of type Ib and Ic with respect to type II
supernovae, may indicate a burst of star formation of young age
($\tau \simlt$ 20 Myr),  a high incidence of binary systems in the
inner regions (r $\simlt 0.4 $R$_{25}$) of Seyfert galaxies, or a
top-loaded mass function.


\end{abstract}

\begin{keywords} galaxies: nuclei - galaxies: Seyfert -- galaxies:
starburst -- stars: formation -- galaxies: stellar content -- stars:
supernov\ae: general
\end{keywords}

\section{Introduction}

The relationship between circum-nuclear star formation and
non-thermal nuclear activity is as yet poorly understood.  While the
possibility that non-thermal activity may be entirely ascribed to
massive stars and to supernova events \cite{terle88,terle94} is
challenged by several lines of evidence, it is unlikely that the
onset of nuclear activity and strong nuclear or circum-nuclear star
formation can be fully unrelated phenomena. Supernov\ae\ may not be
responsible for the optical spectrum and emission line of Seyfert 1
galaxies, but reprocessed gas of supernova ejections may eventually
be accreted by the central black hole, influencing the opacity and
hence the radiating properties of the accretion disk which is
reputed to be one of the ultimate components of the AGN central
engine.

Core-collapse supernov\ae\ from massive progenitors, albeit rare
events, are diagnostics of recent star formation. The  aim of this
paper is to estimate the ratio  between the total number of Ib-Ic
\se\ type II \se\ [\f=N(Ib/c)/N(II)] discovered in a suitable sample
of non-active (hereafter normal) galaxies, as well as to compare it
to that of Seyfert galaxies. As we will show through this paper, the
ratio \f\ reflects metallicity, age, fraction of binary systems, and
initial mass function (IMF) shape effects which are probably much
different in normal and (at least in some) Seyfert galaxies.


\section{Sample Selection \& Criterion \label{sample}}

We cross correlated the latest available Asiago Supernova Catalogue
(last supernova recorded: 2000di on Aug. 23) \cite{barb93} with the
 9$^{\rm th}$\ edition of A Catalogue of Quasars and Active
Galactic Nuclei by M. - P. V\'eron-Cetty and P. V\'eron
\cite{veron2000}. All AGN hosting a supernova have $v_r \simlt
14,000$\kms.  A temporal restriction is needed since type Ib and Ic
\se, first observed in the 1960s \cite{bert64} were labeled as
peculiar ``type I'' \se\ for more than two decades. The realization
that they are a different class of objects came after two decades,
around 1985 \cite{elias85,pan86,pf87}. We limit our counts to \se\
discovered after Dec. 31, 1985. Only few identification of type Ib/c
\se\ discovered earlier than Jan. 1, 1986 are available (as a
product of subsequent re-identification of older type I spectra).

We  define a   sample   of ``normal'' galaxies as including all
galaxies listed in the Asiago Catalogue with $v_r \simlt
14,000$\kms, not classified as Seyfert (including uncertain Seyfert
types, coded as ``S''), \hii\ (code H2) or LINERs (code S3) in the
V\'eron-Cetty and V\'eron lists. The Seyfert sample comprises all
Seyfert galaxies as above but coded as S1, S2 or S1$x$\ in the
catalogue (see Table 1 for sample sizes).

\begin{table}
\begin{minipage}{8cm}
\caption{Distribution of of Supernov\ae\ \label{tab:summ}}
\begin{tabular}{lcccc}
\hline\hline SN Type & \multicolumn{4}{c}{Number of Supernova
Events$^\natural$}
\\
\\
\cline{2-5} & All$^\flat$ &&
\multicolumn{2}{c}{v$_r\le$14000}   \\
\cline{4-5}
 & && Normal & Seyfert
 \\
\cline{2-5}
\\
Unclassified  &       661 &&      36  &  1   \\
I             &        77 &&       3   &  1   \\
Ia            &       565 &&     192   & 14    \\
Ib-Ic         &        81 &&      48    &  6    \\
II            &       374 &&     210   &  6     \\
$S^\sharp$    &   {\em 0.216} && {\em 0.229} & {\em 1.000}\\
Pec           &         6 &&       0   &  0  \\
Total         &      1764 &&     489    & 28 \\
\hline \hline
\end{tabular}

$^\natural$In the time lapse from Jan. 1, 1986 to Aug. 23, 2000 \\
$^\flat$All \se\ listed in the Asiago catalogue, since 1885 until
August 23,
2000.\\
$^\sharp$ f= N(Ib/c)/N(II), as reported in the two preceding rows.
\end{minipage}
\end{table}

We restrict our comparison to the rate of type Ib-Ic and II \se,
since {\em the control time $\tau_c$\ is approximately the same for
both type Ib/c and type II \se}. The difference in brightness and
light curve shape between type Ib-Ic and type II \se\ would
translate in a $\Delta \tau_c/\tau_c \approx$ 15 \%\ \cite{capp93},
which is marginal with respect to the magnitude of the effects we are
considering (a factor $\approx$ 4; see \S \ref{resul}). Even if
Seyfert galaxies and normal galaxies have been monitored over
different times, and with different techniques,  the ratio
N(Ib/c)/N(II) should {\em almost} reproduce the true ratio of  \sa\
rates for type Ib/c and type II (but clearly not each \sa\ rate in
\sa\ units!). In other words, we have SNR(Ib/c)/SNR(II) =
 N(Ib/c)/(0.01 $\tau_{Ib/c}$) $\times$ 0.01
$\tau_{II}$/N(II) $\approx$ N(Ib/c)/N(II), where $\tau$\ is in
centuries,  SNR  is expressed in number of \se\ per year, and
$\tau_c$(II)$\approx\tau_c$(Ib/c).  Selection effects, described in
the next section, will operate in the same way in both galaxy
samples.

The Seyfert sample is dominated by Seyfert 2 galaxies, and at any
rate by very low luminosity AGN. The average absolute magnitude of
the Seyfert  2 sample is $M_B \approx -19.1$ (taking the data of the
V\'eron-Cetty and V\'eron Catalogue)  The S1$x$\ nuclei are, on
average, of similar luminosity ($\overline{M_B} \approx -19.3$). In
complete samples, Seyfert 1 nuclei are on average a factor 100
brighter than Seyfert 2 (see also \S \ref{concl}).

\begin{figure}
\centerline{\psfig{figure=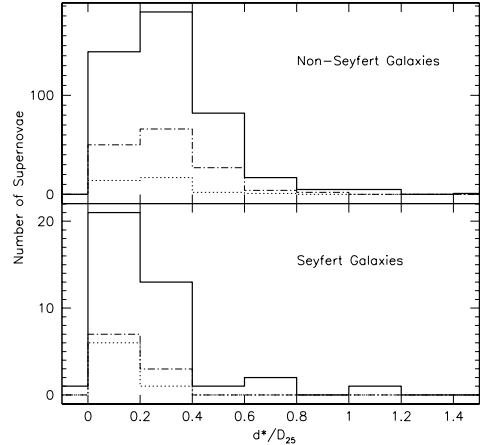,height=7truecm,angle=0}}
\caption{Distribution of supernova distance normalized to the
photometric diameter $\rm D_{25}$, for normal (upper panel) and
Seyfert (lower panel) galaxies with at least one supernova listed in
the Asiago supernov\ae\ catalogue. Supernova offsets and photometric
diameters are as reported in the Asiago catalogue. Abscissa scale is
ratio between deprojected supernova distance and photometric radius;
ordinate scale is number of supernov\ae. Thick solid line: all \se;
dotted lines: type Ib/c \se, dot-dashed line: type II \se.
}\label{fig:distnorm}
\end{figure}

\begin{figure}
\centerline{\psfig{figure=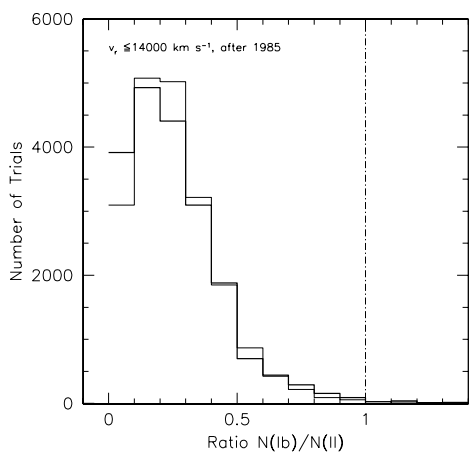,height=7.0truecm,angle=0}}
\caption{Distribution of the ratio \f = N(Ib/Ic)/N(II) for 20000
bootstrap replications of the Seyfert sample by pseudo-control
samples of the same size (see text). Thick solid lines: without
redistribution of unknown and type I \se; thin solid line: after
redistribition of unknown and type I \se. We considered 25 galaxies
with $v_r \le 14000$\kms\ with \se\ discovered in or after 1985. The
vertical dot-dashed line marks the value found for the Seyfert
sample, as reported in Table 1. }\label{fig02}
\end{figure}

\begin{figure}
\centerline{\psfig{figure=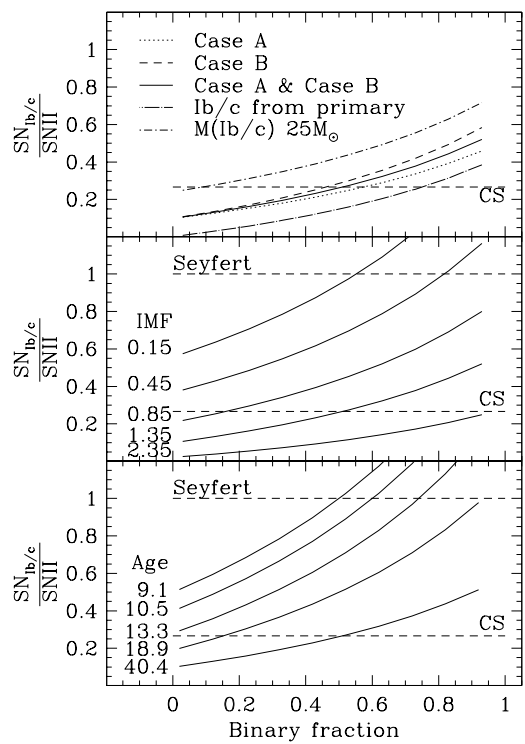,width=8.0truecm,angle=0}}
\caption{ Upper panel: expected ratio \f\ of SNIbc to SNII rates as
a function of the binary fraction. A Salpeter IMF and a constant
star formation rate in the last 100 Myr have been adopted. Several
cases are considered as detailed in the text. Middle panel: IMF
effects. Ratio \f\ in model c), for different IMF exponents x, as
labelled in the figure. The Salpeter IMF corresponds to x=1.35.
Lower panel: age effects. Ratio \f\ in model c) for different
starburst ages (assuming solar composition), as  labelled in Myr.}
 \label{fig:bin}
 \end{figure}

\section{Results }\label{resul}

\subsection{The Distribution of Core Collapse Supernova Types}

Table 1 reports the number of \se\ counted in each sample, with
the restrictions on redshift and on discovery time motivated
above.


The ratio  \f\ = N(Ib--Ic)/N(II) for normal galaxies is
$\approx$0.23, very close to the values obtained from the ratio of
absolute \sa\ rates in SNu: 0.18 for  S0a-Sb galaxies, and 0.27 for
Sbc-Sd galaxies \cite{capp97}.  The ratio \f\ is much larger in the
Seyfert than in the normal galaxies samples, by a factor $\approx 4$.
 \f $_{cc,Sy} = N(Ib/Ic)/(N_{cc}, \approx 0.5 \gg$
\f$_{norm} \approx 0.19$, where N$_{cc}$ = N(Ib) + N(Ic) + N(II) is
the number of core-collapse \se. Assumption of binomial statistics,
where the probability is p=\f $_{cc,norm}$, yields an excess of type
Ib-Ic supernova in the Seyfert sample significant to a confidence
level of $\ga$ 2$\sigma$ (N(Ib/c) expected in the Seyfert sample is
2.3$\pm$1.3). However, this simple estimate may not be fully
satisfactory, since normal galaxies and Seyfert samples may be
influenced by factors affecting the probability of discovery as well
as the intrinsic rate of \sa\ explosion (see \S \ref{stat}).

\subsection{The Radial Distribution of Supernov\ae}

Fig. \ref{fig:distnorm} shows the distribution of the \sa\
de-projected distance d$^*$\  normalized to the standard isophotal
diameter D$_{25}$, for normal and Seyfert galaxies, computed under
the assumptions that all galaxies can be considered flat disks.
Supernov\ae\ in Seyfert galaxies show a tendency to occur closer to
the nucleus.  A Kolmogorov -- Smirnov test suggests that the
distributions are different to a statistical significance of 0.999.
This result has been known for some time \cite{petr90,kaz97} from
the analysis of \sa\ data in different samples.

\subsection{Significance of Results: Is there any Real Excess of type
Ib-Ic Supernovae in Seyfert galaxies? \label{stat}}

The ratio \f$_{norm}$ has been determined from 210 type II and 48
type Ib-Ic supernovae; therefore it is possible to safely use the
binomial statistics to write \f$_{norm} \approx$ 0.23 $\pm$ 0.05 (at
1 $\sigma$ confidence level). To properly estimate the statistical
significance of the excess of type Ib-Ic  supernov\ae\ in Seyfert
galaxies, we used a ``bootstrap'' method.  We generated a large
number ($\sim 10^4$) of synthetic data sets matching the Seyfert
galaxy sample. The synthetic data set were extracted randomly from
our normal galaxies samples, with the same number of galaxies as in
the Seyfert samples, and with the same {\em distribution} of any
parameter influencing either the probability of discovery of \se, or
the intrinsic rate of supernova production \cite{capp93}: (1) galaxy
inclination, (2) D$_{25}$, (3) morphological type, (4) radial
velocity. ``Matching'' samples were considered all synthetic samples
for which the four distributions were {\em all} not statistically
different from the distributions for the  Seyfert sample at a
$2\sigma$\ confidence level. In other words, since we could not
eliminate selection effects from our Seyfert sample, we reproduced
them in a similar fashion in a large number of ``pseudo-control
samples.''

In all cases save the morphological type,  the statistical difference
between the parameter distribution  of the Seyfert  galaxy sample
and of same-size samples of normal galaxies was assessed by  a
Kolmogorov-Smirnov test. In the case of morphological type, a
$\chi^2$\ test was applied, after rebinning of T values into four
bins for early type galaxies (E-S0), early spirals (S0a-Sb)  late
type galaxies (Sc and later), and unknown or peculiar. We computed
the number of \se\ for each synthetic data set, and then we counted
which fraction of synthetic data sets had \sa\ numbers exceeding the
number from the real data. That fraction has been assumed to be our
statistical significance \cite{efron93}.


Fig. \ref{fig02} the thick solid line shows the distribution of \f\
for the pseudo-control samples. The dot-dashed line marks the value
found for the Seyfert sample. Only in $\approx$ 1 \%\ of the cases a
larger value of the ratio N(Ib/c)/N(II) is found, confirming that a
realistic estimate of the significance of the result is $\ga$
2$\sigma$. The thin line marks the distribution of \se\ after \se\
whose classification was unknown and or of generic type ''I'' have
been re-assigned (consistently with the approach of \cite{petr90}).
From Fig. \ref{fig02}, we infer that the redistribution for
unclassified \sa\ types has a negligible effect on the statistical
results.


\section{Inferences on Star Formation \label{disc}}

\subsection{Normal Galaxies}

The interpretation of the ratio \f\ for the ``average normal" galaxy is
straightforward, if we consider most likely that Ib/c  \se\
progenitors are Wolf Rayet stars, whose hydrogen envelope has
been fully lost \cite[and references therein]{filippo98rev}. The
latter stars may result from the evolution of the most massive
stars (M$\geq$40M$\odot$)\cite{bress81,bress94}
or from less massive primary component of binary systems
after the mass transfer (eg. Table 2 of \cite{well99}).
Assuming
a constant star formation rate during the last 100~Myr, and
adopting the same initial mass function (IMF) for single stars and
binary components,
the ratio \f\ is simply given by
\begin{equation}
S=\frac{\sum_iW^i\int_{m_1^i}^{m_2^i}\psi(m)dm}{\sum_jW^j\int_{m_1^j}^{m_2^j}\psi(m)dm}
\label{eq:f}
\end{equation}
where m is the initial mass of the star, the index i refers to all
channels leading to type Ib/c \se, j to the formation of type II
\se\ and W's are relative weights that depend on the overall binary
fraction and on the relative frequency of case A, B and C in the
binary evolution.

The expected variation of the ratio \f\ is shown in the upper panel
of Fig.\ref{fig:bin}, as a function of the binary fraction. To this
purpose we have adopted a Salpeter IMF up to 120~M$\odot$, and the
values of m$^i$ and m$^j$ were taken from Table 2 of \cite{well99};
the latter authors also assume that all stars more massive than
40~M$\odot$ explode as type Ib/c \se, while  single stars and
secondary stars less massive than 40~M$\odot$ explode as type II \se.
We have considered several cases: (a) all binaries evolve as Case A
(dotted line); (b)  all binaries evolve as Case B (dashed line); (c)
Case A and B share the same relative weight (solid line); (d) as in
(c) but in the case that only binary members explode as type Ib/c
\se\ (lower dot long-dashed line); (e) as in (c) but in this case
the initial mass for Ib/c \se\ has been lowered from 40~M$\odot$ to
25~M$\odot$ (upper dot short-dashed line). The horizontal line
represents the observed average value for the normal galaxy sample
(CS). We did not account for Case C evolution because it is
considered to be a very rare event and as such it will only slightly
modify the picture depicted in Fig.\ref{fig:bin}. The fraction \f\
does not appear much affected by the precise details of the binary
evolution. For normal spirals, the observed ratio is consistent with
a scenario where  the progenitors of type Ib/c \se\ are Wolf Rayet
stars, either evolved from stars initially more massive than
40~M$_\odot$ or from less massive primary components of binary
systems. The data support  a universal Salpeter IMF and indicate a
binary fraction of about the 50\%. A significantly higher binary
fraction (about 75\%) is required if one insists that only binary
evolution give rise to type Ib/c \se. As discussed by Wellstein \&
Langer (1999), uncertainties in the mass- loss rate, convection
treatment and chemical composition may affect both the upper mass
limit (40~M$\odot$) leading  to type II \se\ in single stars, and
the mass transfer phase in binary evolution. Fortunately,
uncertainties leading to different type Ib and type Ic \se\ events
are not relevant here, because we are considering their combined
frequency.

\subsection{Seyfert Galaxies}

From Table 1 and Fig.\ref{fig:bin}, we also infer that things could
be markedly different in Seyfert galaxies, where the observed ratio
is \f$\approx$1. Several Seyfert galaxies, of which a sizable
fraction belongs to strongly  interacting systems \cite[$\approx$
2/3 of Seyfert 2]{dultzin97,dultzin99}, show significant, or even
dominant (over the active nucleus luminosity) circum-nuclear star
formation. A recent \ha\ narrow band survey \cite{g-d01} found that
about 1/2 of Seyfert 1 and Seyfert 2 show maximum value of \hii\
region emission in the circum-nuclear regions. This study of the
radial distribution of \hii\ regions in normal, LINER and Seyfert
galaxies agrees with a large number of detailed spectroscopic and
imaging studies of Seyfert galaxies. Consistently, the radial
distribution of \se\ appears to be more strongly peaked in Seyfert
galaxies than in normal galaxies in several samples
\cite{kaz97,petr90} as well as in our own. However, enhancement of
star formation rate  in the inner galactic regions cannot explain
the relative excess of type Ib/c \se\  found in Seyfert galaxies.

The observed radial gradients of chemical composition \cite{vce92}
and the central concentration of type Ib/c \se\ in Seyfert galaxies
(Fig. \ref{fig:distnorm}) imply that Ib/c \se\ could be typically
produced in 2 $Z_\odot$ regions in Seyfert and in $Z_\odot$\ regions
in normal galaxies. Because  mass loss increases with metallicity
and, consequently, the minimum initial mass for single Wolf Rayet
production decreases, \f\ could be a function of metallicity.  As
shown by model (d) in the upper panel of Fig.\ref{fig:bin}, \f\
could increase by the 100\%\, if the minimum mass for Ib/c
production decreases from 40 M$_\odot$ to 25 M$_\odot$. However,
this is not very likely because mass-loss is also sensitive to the
luminosity and the latter is a strong function of the initial mass
\cite{bress94}.

Another possible explanation is that star formation in
circum-nuclear and disk regions of Seyfert galaxies is a
 high {\em binary fraction}. If we assume that all
young stars are in binaries and undergo mass transfer during their
life, we would obtain \f$\simeq$0.6. This value, though lower than
the observed one, would double the expected relative frequency of
binary stars with respect to the normal galaxies. Inspection of
Table 1 shows that  the observed ratio N(Ia)/N$_{cc}$\  is 0.63 for
the normal galaxies and rises to 1 in the Seyfert sample. Would this
ratio be free from selection effects, it would strongly indicate a
larger probability of assembling binary systems. However,  the
magnitude at maximum and light-curve of type Ia \se\ is very
different from that of core-collapse \se. This makes  the comparison
of the ratio N(Ia) over number of core-collapse \se\ {\em
meaningless} if not accompanied by a careful examination of the
relative control times which, unfortunately cannot be done here.

An appealing alternative is constituted by a ``top-loaded'' IMF. The
high efficiency at which gas is reprocessed into stars is often
taken as an indication in favor of a top-heavy initial mass function
in compact Starbursts e.g. \cite{franceschool}. The middle panel of
Fig. \ref{fig:bin} depicts the effect of changing the IMF exponent
x, where $\psi(m)\propto{m^{-(1+x)}}$ and x=1.35 corresponds to the
Salpeter IMF. The value of  \f\  increases as the IMF becomes
flatter. The observed value in Seyfert galaxies would indicate a
very flat IMF, above 8 $M_\odot$. We must however stress that
conclusive evidence in favor of an IMF slope significantly flatter
than the Salpeter one has not been  found in external galaxies, not
even in Wolf-Rayet and Starburst galaxies \cite{sch99,leit99}.

The necessity of a top heavy IMF  disappears if, in Seyferts, we are
observing a very young burst of star formation. To illustrate this
effect we show in the lower panel of Fig.\ref{fig:bin} the run of
the ratio \f\ with the binary fraction, for different burst ages and
a constant SFR. If the starburst is sufficiently old, let us say of
age $\tau^\star \simgt$ 50 Myr, then there may exist evolved stars
with any mass between the upper IMF limit and  the critical mass
M$\simeq$8~$M_\odot$ for core collapse \se. But if the burst is
younger, then all stars with initial mass below a given value --
M($\tau$) -- are still on the Main Sequence and only stars more
massive than M($\tau^\star$) can explode. Different curves in the
lower panel of Fig. \ref{fig:bin} are labeled by the duration of the
burst, in Myr. A Salpeter IMF is consistent with the observed
\f$_{Seyf}$\ if the star formation episode is younger than $\approx$
13 Myr. This value depends only slightly on the assumed chemical
composition.

\section{Conclusions }\label{concl}

We found that normal galaxies show \f$_{norm} \approx $0.27, a value
consistent with the ratio of absolute supernova rates in SNu. With
the caveat of our still-limited sample size,  Seyfert galaxies show
a peculiar distribution of supernova types, with higher frequency of
type Ib/c \se\ (\f$_{Seyf} \approx~1$) than non-active galaxies.
These finding are consistent with a ``normal'' \sa\ rate related to
secular star formation as far as non-active galaxies are concerned.
A scenario where Wolf Rayet stars are produced by the most massive
stars ($\geq$40M$_\odot$) and by less massive
(8M$_\odot$$\leq$M$\leq$40M$_\odot$) primary stars of binary systems
that soon or later undergo mass-transfer, reproduces fairly well the
observed ratio, without particular assumptions on the IMF.

A large type Ib/c rate with respect to type II \se, as found for
Seyferts, is a new result. We consider several explanations as
possible, namely a lowering of the upper mass limit for type II
precursors due to a higher metallicity, a very high {\em binary
fraction}, a top heavy IMF and a young age of the burst of star
formation. While metallicity is likely to act in the observed
direction, because it enhances the mass-loss rate, the amplitude of
the observed effect  may not be explained.

Even if the \f\ value found for Seyfert galaxies  needs confirmation
with better statistics, it is interesting to note that an high \f\
could be associated with the enhanced SFR in the circum-nuclear and
regions of galaxies hosting Seyfert nuclei, especially of type 2
nuclei \cite{dd95}. A young age of the burst of star formation --
probably triggered by interaction with a massive companion galaxy
\cite{dultzin99} -- is a likely cause of the high \f\ observed among
Seyfert galaxies, as it requires the minor number of assumptions.
This interpretation is also consistent with an evolutionary sequence
leading from IR-luminous Starbursts, to Seyfert 2 and eventually to
Seyfert 1 galaxies, with the youngest Seyfert 1 being the ones of
lowest intrinsic power and/or the ones most heavily obscured
\cite{gu01}.


\bigskip
AB thanks the {\sl Osservatorio di Collurania} at Teramo (Italy)  for
their hospitality. Two of us  acknowledge financial support by the
Italian Ministry for University and Research (MURST) under grant
Cofin
 00-02-007 (PM) and Cofin 9902763542\_004 (AB).





\begin{thebibliography}{}
\bibitem[\protect\citename{Barbon et al. }1993]{barb93} Barbon, R., Benetti, S., Cappellaro, E., Patat, F.,
\&\ Turatto, M. 1993, The Asiago Supernova data base, Memorie della
Societa Astronomica Italiana, 64, 1083
\bibitem[\protect\citename{Bertola }1964]{bert64} Bertola, F.\ 1964, AJ, 69,
236
\bibitem[\protect\citename{Bressan, Chiosi, \& Bertelli }1981]{bress81} Bressan,
A.\ G., Chiosi, C., \& Bertelli, G.\ 1981, A\&A, 102, 25
\bibitem[\protect\citename{Bressan }1994]{bress94} Bressan, A. 1994, Space Science
Reviews, 66, 373
\bibitem[\protect\citename{Bressan, Chiosi, \&\ Fagotto  }1994]{bressetal94} Bressan,
A., Chiosi, C., \& Fagotto, F.\ 1994, ApJS, 94, 63
\bibitem[\protect\citename{Cappellaro et al. }1993]{capp93} Cappellaro, E., Turatto, M.,
Benetti, S., Tsvetkov, D. Yu., Bartunov, O. S., Makarova, I. N.
1993, A\&A, 268, 472
\bibitem[\protect\citename{Cappellaro et al. }1997]{capp97} Cappellaro, E.,
Turatto, M., Tsvetkov, D.Y.., Bartunov, O.S., Pollas, C., Evans, R.,
\&\ Hamuy, M. 1997,  A\&A, 322, 431
\bibitem[Dultzin-Hacyan(1995)]{dd95} Dultzin-Hacyan, D.\
1995, RMxAA Conference Series, 3, 31
\bibitem[\protect\citename{Dultzin-Hacyan et al. }1999]{dultzin99} Dultzin-Hacyan, D., Krongold, Y.,
Fuentes-Guridi, I., \&\ Marziani, P. 1999, ApJ, 513, L111
\bibitem[\protect\citename{Efron \&\ Tibshirani }1993]{efron93} Efron, B. \&\ Tibshirani, R. J.
1993, An Introduction to the Bootstrap (New York: Chapman \&\
Hall)
\bibitem[\protect\citename{Elias et al. }1985]{elias85} Elias, J.~H., Matthews, K.,
Neugebauer, G., \& Persson, S.~E.\ 1985, ApJ, 296, 379
\bibitem[\protect\citename{Filippenko }1998]{filippo98rev} Filippenko, A.   V. 1998, ARA\&A, 35,
309
\bibitem[Franceschini 1999]{franceschool} Franceschini, A. 1999,
The Far-Infrared and Sub-Millimeter View, in
"Galaxies at High Redshift", Tenerife November 1999, I.
Perez-Fournon, M. Balcells, F. Moreno-Insertis and F. Sanchez Eds.,
Cambridge University Press
\bibitem[\protect\citename{Gonz{\' a}lez-Delgado, Heckman, \& Leitherer }2001]{g-d01} Gonz{\' a}lez Delgado, R.\ M., Heckman, T.,
\& Leitherer, C.\ 2001, ApJ, 546, 845
\bibitem[\protect\citename{Gu, Maiolino, \&\ Dultzin-Hacyan }2001]{gu01} Gu, Q., Maiolino, R., \& Dultzin-Hacyan, D.\ 2001, A\&A, 366, 765
\bibitem[\protect\citename{Kazarian }1997]{kaz97} Kazarian, M. A.  1997, Astrophysics, 40,  296
\bibitem[Leitherer(1999)]{leit99} Leitherer, C.\ 1999, IAU
Symp.~186: Galaxy Interactions at Low and High Redshift, 186, 243
\bibitem[\protect\citename{Panagia, Sramek, \& Weiler }1986]{pan86} Panagia,
N., Sramek, R.\ A., \& Weiler, K.\ W.\ 1986, ApJ, 300, L55
\bibitem[\protect\citename{Petrosian \&\ Turatto }1990]{petr90} Petrosian, A. R., \&\ Turatto, M.
1990, A\&A 239, 63
\bibitem[\protect\citename{Porter \& Filippenko  }1987]{pf87} Porter, A.\ C.\
\& Filippenko, A.\ V.\ 1987, AJ, 93, 1372
\bibitem[\protect\citename{Richmond, Filippenko, \& Galisky }1998]{filippo98}
Richmond, M. W., Filippenko, A. V., \&\ Galisky, J. 1998,  PASP, 110,
553
\bibitem[\protect\citename{Schaerer, Contini, \& Kunth }1999]{sch99} Schaerer,
D., Contini, T., \& Kunth, D.\ 1999, A\&A, 341, 399
\bibitem[\protect\citename{Terlevich et al. } 1993]{terle93} Terlevich, R. J.,
Tenorio-Tagle, G., Franco, J., Rozycska, M., \&\ Boyle, B. J. 1993,
RMxAA, 27, 59
\bibitem[\protect\citename{Terlevich \&\ Melnick }1988]{terle88} Terlevich, R. \&\
Melnick, J. 1988, Nature, 333, 239
\bibitem[\protect\citename{V\'eron-Cetty \&\ V\'eron}2000]{veron2000} Veron, P., \&\
Veron-Cetty, M. - P. 2000, A Catalogue of Quasars and Active Nuclei,
9$\rm ^{th}$ Edition. ESO Scientific Report No. 19
\bibitem[\protect\citename{Vila-Costas \& Edmunds }1992]{vce92}
Vila-Costas, M. B., \& Edmunds, M. G. 1992, MNRAS, 259, 121
\bibitem[\protect\citename{Wellstein \& Langer }1999]{well99} Wellstein, S.\ \&
Langer, N.\ 1999, A\&A, 350, 148

\end{thebibliography}
\end{document}